\PassOptionsToPackage{table,dvipsnames,svgnames}{xcolor}
\PassOptionsToPackage{hyphens}{url}
\documentclass[sigconf, screen, nonacm]{acmart}

\AtBeginDocument{%
  }

\usepackage{tikz}
\usetikzlibrary{positioning,arrows.meta}
\usepackage{booktabs}  % For professional-looking horiz lines (\toprule, \midrule, \bottomrule)
\usepackage{paralist}

\usepackage[utf8]{inputenc}    
\usepackage{enumitem}    
\usepackage{balance}
\usepackage{comment}
\usepackage{subcaption}
\usepackage{adjustbox}
\usepackage{tabularx}      % For tables with columns that automatically adjust width
\usepackage{longtable}     % For multipage tables (used in appendix raw data)
\usepackage{makecell}      % For multi-line cells, used here for the headers
\usepackage{textcomp}      % For symbols like the registered trademark symbol
\usepackage{ragged2e}      % For better text alignment (e.g., \RaggedRight)
\usepackage{threeparttable}
\usepackage{cleveref}
\crefname{algocf}{algorithm}{algorithms}
\Crefname{algocf}{Algorithm}{Algorithms}
\usepackage[T1]{fontenc} 
\usepackage{iftex} % to access the \ifLuaTeX command
\newcolumntype{L}{>{\RaggedRight\arraybackslash}X}
\newcolumntype{C}{>{\centering\arraybackslash}X}

\definecolor{tablehead}{HTML}{442266}

\newcommand{\param}[1]{\textcolor{red}{#1}} % to highlight hardcoded numbers
\newcommand{\xxx}[1]{\param{XXX}} % to highlight hardcoded numbers

\newcommand{\ignore}[1]{}

\newcounter{obs}
\definecolor{nbs}{rgb}{0.88, 0.07, 0.37}
\definecolor{agyc}{rgb}{0.37, 0.88, 0.07}
\definecolor{moegi}{rgb}{0.357, 0.537, 0.188}
\definecolor{burntorange}{rgb}{0.8, 0.33, 0.0}
\definecolor{carmine}{rgb}{0.59, 0.0, 0.09}
\definecolor{ceruleanblue}{rgb}{0.16, 0.32, 0.75}

\newif\ifcamerareadyiterations
\camerareadyiterationsfalse

\newif\ifcameraready
\camerareadytrue

\newif\ifdraft
\draftfalse

\newif\ifblind
\blindtrue

\newif\ifispassdraft
\ispassdraftfalse

\newif\ifshepherd
\shepherdfalse

\ifshepherd
\usepackage[colorinlistoftodos,prependcaption,textsize=small]{todonotes}

\newcommand{\scomment}[2]{\todo[size=\scriptsize, linecolor=ceruleanblue, bordercolor=ceruleanblue, backgroundcolor=white!90!ceruleanblue]{\textcolor{ceruleanblue}{
\textbf{S#1} #2}}}
\newcommand{\rcomment}[1]{\todo[size=\scriptsize, linecolor=burntorange, bordercolor=burntorange, backgroundcolor=white!90!burntorange]{\textcolor{burntorange}{
\textbf{#1}}}}

\else

\newcommand{\scomment}[2]{}
\newcommand{\rcomment}[1]{}

\fi

\ifdraft
    \usepackage[colorinlistoftodos,prependcaption,textsize=small]{todonotes}
    \newcommand{\nbcomment}[1]{\todo[size=\scriptsize, linecolor=orange, bordercolor=orange, backgroundcolor=white]{\textcolor{nbs}{\textbf{@nb:} #1}}}

    \newcommand{\mma}[1]{\textcolor{orange}{\textbf{[@mma: #1]}}}
    \newcommand\mmatodo[1]{{\textcolor{orange}{#1}}}
    \newcommand{\atbcomment}[1]{\todo[size=\scriptsize, linecolor=orange, bordercolor=orange, backgroundcolor=white]{\textcolor{blue}{\textbf{@atb:} #1}}}

    \newcommand{\omcrcomment}[1]{\todo[size=\scriptsize, linecolor=orange, bordercolor=orange, backgroundcolor=white]{\textcolor{blue}{\textbf{@om:} #1}}}

\else 
    \newcommand{\nbcomment}[1]{}

    \newcommand{\mma}[1]{}
    \newcommand\mmatodo[1]{}
    \newcommand\atbcomment[1]{}
    \newcommand\mmacomment[1]{}

    \newcommand{\omcrcomment}[1]{}
\fi

\ifispassdraft
    \usepackage[colorinlistoftodos,prependcaption,textsize=small]{todonotes}
    \renewcommand{\nbcomment}[1]{\todo[size=\scriptsize, linecolor=orange, bordercolor=orange, backgroundcolor=white]{\textcolor{nbs}{\textbf{@nb:} #1}}}
    
\fi

\definecolor{aliceblue}{rgb}{0.94, 0.97, 1.0}

\usepackage[framemethod=tikz]{mdframed}

\usepackage[most]{tcolorbox} 
\newtcolorbox[auto counter]{tkx}[2][]{%
    enhanced, breakable, center title,
    colframe = #2!45,
    colback  = #2!10,
    colbacktitle=#2!20,
    left=1.5pt,
    right=1.5pt,
    bottom=1.5pt,
    top=1.5pt,
    #1% 
}

\newif\ifextendedappendix
\extendedappendixfalse
\extendedappendixtrue

\hypersetup{
    colorlinks=true,
    linkcolor=ForestGreen,
    citecolor=ForestGreen,
    urlcolor=RoyalBlue,
    filecolor=IndianRed,
    pdftitle={Control-Flow Profiling for Workload Representativeness},
    pdfauthor={Mahesh Madhav},
    pdfsubject={Benchmarks},
    pdfkeywords={Dynamic binary instrumentation, control-flow profiling, workload representativeness, benchmarking, Valgrind, basic-block vectors},
    pdfnewwindow=true,
    pdfdisplaydoctitle=true,
    bookmarksopen=true
}
\title{Control-Flow Profiling for Workload Representativeness}

\author{Mahesh Madhav}
\affiliation[obeypunctuation=true]{%
  \institution{Ampere Computing}, \city{Portland}, \state{OR}, \country{USA}
}
\email{mahesh@amperecomputing.com}
\hypersetup{pdfauthor={Mahesh Madhav}}

\begin{document}

\begin{abstract}

Workload characterization in performance engineering often replaces expensive target executions with lower-cost proxies, including reduced datasets, miniapps, and training inputs. Such substitutions require empirical verification that the proxy faithfully exercises the target's dynamic execution behavior. This paper presents an address-keyed control-flow profiling methodology and tool suite designed to evaluate workload representativeness directly from compiled, unmodified binaries. We pair time-ordered basic-block vector (BBV) recurrence analysis with \texttt{exp-branchprof}, an experimental Valgrind tool that captures five orthogonal structural profile families spanning branch direction bias, four-bit global history contexts, dynamic indirect targets, function call transitions, and intraprocedural CFG edges.

Our method compares weighted event distributions from two executions of the same binary.  It reports support coverage, reference event-mass coverage, distribution distance, entropy-derived quantities, and phase structure, rather than claiming that a single similarity score certifies a proxy.  We validate the tool with exact-oracle microbenchmarks, covering conditional and inverted exits, calls, returns, direct and indirect jumps, and forked processes.  We then apply the workflow to SPEC CPU2017 and CPU2026 train/reference pairs and seven refrate/refspeed pairs.  The corpus contains both close matches and cases where a shorter execution reaches much of the same code but misses substantial target, callgraph, CFG, or BBV event mass.  The accompanying artifact supports the same analysis for proprietary workload/proxy pairs without requiring raw PC-level profiles to leave the collection environment.
\end{abstract}

\maketitle
\pagestyle{plain}
\hypersetup{pdfauthor={Mahesh Madhav}}
\newcommand{\Red}[1]{{\color{red} #1}}
\newcommand{\asm}[1]{\texttt{#1}}
\newcommand{\sys}[1]{\texttt{#1}}
\newcommand{\kw}[1]{\textit{#1}}
\newcommand{\kwb}[1]{\textbf{#1}}
\newcommand{\type}[1]{\textit{#1}}
\newcommand{\Response}[1]{{\color{blue} #1}}
\newcommand{\XXX}[1]{\Red{\textbf{XXX[}#1\textbf{]}}}
\newcommand{\tocite}[1]{\Red{CITE:\cite{#1}}}
\newcommand{\toref}[1]{\Red{REF:\ref{#1}}}
\newcommand{\parasub}[1]{\smallskip\noindent\textit{{#1:}\xspace}}
\newcommand{\anyone}[1]{\textcolor{blue}{#1}}
\newcommand{\niparagraph}[1]{\noindent\textbf{\textsf{#1}\hspace{0.5em}}}
\newcommand\TODO[1]{\textcolor{red}{TODO: #1}}

\newenvironment{CompactItemize}%
  {\begin{list}{$\blacktriangleright$}%
    {\leftmargin=\parindent \itemsep=2pt \topsep=2pt
     \parsep=0pt \partopsep=0pt}}%
  {\end{list}}
\renewcommand{\labelitemi}{$\blacktriangleright$}

\newcommand{\malloc}{{\texttt{Malloc}}}
\newcommand{\linklist}{{\textsf{Linked-List}}}
\newcommand{\plusplus}{\texttt{++}}

% Gradient table macros
% Gradient helpers for profile-fidelity tables.
% Requires: xcolor/colortbl, etoolbox, pgfmath.

\ifdefined\gradient
\else

\definecolor{low}{HTML}{E66101} % orange for deficit from 1.000
\definecolor{mid}{HTML}{FFFFFF}
\definecolor{high}{HTML}{FFFFFF}
\definecolor{keylow}{HTML}{FFFFFF}
\definecolor{keymid}{HTML}{E2F0D9}
\definecolor{keyhigh}{HTML}{70AD47}
\newcommand*{\opacity}{90}

\newcommand*{\minval}{0.100}
\newcommand*{\midval}{0.900}
\newcommand*{\maxval}{1.001}

% Forward gradient: low values are worse and become orange (deficit from 1.000).
\newcommand{\gradient}[1]{
    \ifdimcomp{#1pt}{>}{\maxval pt}{#1}{
        \ifdimcomp{#1pt}{<}{\minval pt}{#1}{
            \ifdimcomp{#1pt}{<}{\midval pt}{
                \pgfmathparse{int(round(100*(#1-\minval)/(\midval-\minval)))}
                \xdef\tempa{\pgfmathresult}
                \cellcolor{mid!\tempa!low!\opacity} #1
            }{
                \pgfmathparse{int(round(100*(#1-\midval)/(\maxval-\midval)))}
                \xdef\tempa{\pgfmathresult}
                \cellcolor{high!\tempa!mid!\opacity} #1
            }
    }}
}

% Alias for orange gradient
\let\orangegradient\gradient

% Inverted gradient: high values are worse and become red.
% Useful for distances such as total variation distance.
\newcommand{\invgradient}[1]{
    \ifdimcomp{#1pt}{>}{\maxval pt}{#1}{
        \ifdimcomp{#1pt}{<}{\minval pt}{#1}{
            \pgfmathparse{int(round(100*(#1-\minval)/(\midval-\minval)))}
            \xdef\tempa{\pgfmathresult}
            \cellcolor{low!\tempa!mid!\opacity} #1
    }}
}

% Key-count gradient: #1 is a normalized reference-key count in [0,1],
% and #2 is the displayed train/reference KKeys pair.
\newcommand{\keygradient}[2]{
    \ifdimcomp{#1pt}{>}{1pt}{\cellcolor{keyhigh!\opacity} #2}{
        \ifdimcomp{#1pt}{<}{0pt}{#2}{
            \ifdimcomp{#1pt}{<}{0.5pt}{
                \pgfmathparse{int(round(100*#1/0.5))}
                \xdef\tempa{\pgfmathresult}
                \cellcolor{keymid!\tempa!keylow!\opacity} #2
            }{
                \pgfmathparse{int(round(100*(#1-0.5)/0.5))}
                \xdef\tempa{\pgfmathresult}
                \cellcolor{keyhigh!\tempa!keymid!\opacity} #2
            }
    }}
}

% PGO speedup gradient: ratios below 1.0 are worse and become red; ratios
% above 1.0 are better and become green.  The displayed range is [0.65, 1.35].
\newcommand{\pgogradient}[1]{
    \ifdimcomp{#1pt}{>}{1.35pt}{\cellcolor{keyhigh!\opacity} #1}{
        \ifdimcomp{#1pt}{<}{0.65pt}{\cellcolor{low!\opacity} #1}{
            \ifdimcomp{#1pt}{<}{1.0pt}{
                \pgfmathparse{int(round(100*(#1-0.65)/(1.0-0.65)))}
                \xdef\tempa{\pgfmathresult}
                \cellcolor{mid!\tempa!low!\opacity} #1
            }{
                \pgfmathparse{int(round(100*(#1-1.0)/(1.35-1.0)))}
                \xdef\tempa{\pgfmathresult}
                \cellcolor{keyhigh!\tempa!mid!\opacity} #1
            }
    }}
}

\fi

 % aliases

\section{Introduction}
\label{sec:introduction}

CPU performance engineering often relies on substituting a smaller proxy for a
long target workload \cite{heroux_miniapps_2009, aaziz_proxy_correspondence_2018}.  The proxy may be a benchmark train input, a miniapp, a
reduced production trace, or a short simulation region.  The substitution
lowers collection and simulation cost, but it also creates a representativeness
question.  Does the proxy preserve the dynamic behavior that the downstream
decision needs?

\begin{figure}[t]
  \centering
  \includegraphics[width=\columnwidth]{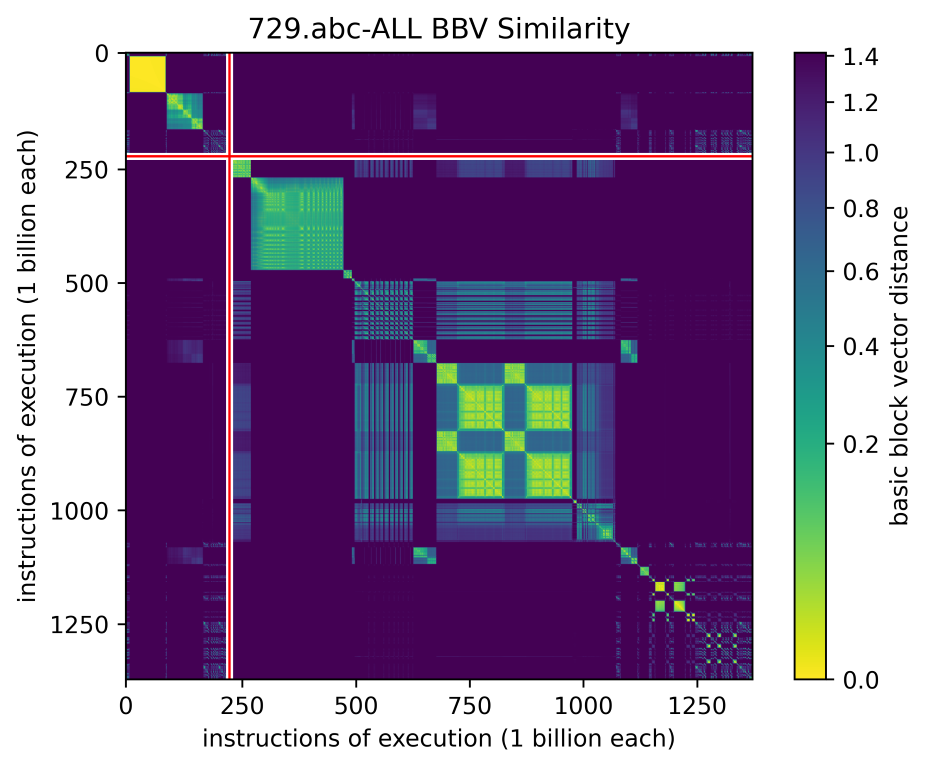}
  \caption{BBV phase self-similarity for \texttt{729.abc} (train proxy
  concatenated with reference target in execution order). The training proxy
  achieves only \(35.7\%\) static block coverage and \(48.8\%\) dynamic
  event-mass coverage, missing over half (\(51.2\%\)) of the reference
  instruction mass and failing to cover major reference phase clusters.}
  \label{fig:abc-bbv}
  \Description{BBV self-similarity recurrence plot for 729.abc showing phase mismatch and unexercised reference event mass.}
\end{figure}

This question is consequential for compiler profiling, simulation, pre-silicon
emulation, and workload selection.  A proxy can reach much of the same code as
its target while changing which blocks, branches, call edges, and CFG edges are
hot.  Code reachability alone therefore cannot establish that a proxy is
appropriate for a frequency-sensitive use.  Earlier work on reduced inputs,
simulation points, and SPEC training workloads established that input choice
can alter observed program behavior
\cite{minnespec, simpoints, bbv_distribution, cpu2006_reftrain,
cpu2006_workshop}.  As illustrated by \texttt{729.abc} in \autoref{fig:abc-bbv},
a proxy can diverge significantly from its target.  The \texttt{abc} train input exercises an
isolated subset of logic-synthesis commands, leaving over half of the
reference workload's dynamic execution unobserved.  We provide a
reusable, binary-level method for making that comparison explicit.

Our design deliberately separates \emph{measurement} from speculative downstream
utility.  A profile can be accurate with respect to one execution yet not
directly predict the behavior of a particular compiler, hardware predictor, or
simulator.  Langdale and Gross emphasize that profile accuracy and optimization
utility are distinct \cite{langdale_profile_usefulness_2003}.  Accordingly,
this paper does not attempt to reduce workload representativeness to a single
speculative prediction score or performance model.  Instead, it provides direct,
multi-view dynamic observability into where two executions agree or diverge.

The method combines two complementary forms of observation.  Basic-Block
Vectors (BBVs) summarize execution by basic-block activity and retain a
time-ordered phase view \cite{simpoints}.  For control flow we introduce
\texttt{exp-branchprof}, built on the Valgrind dynamic binary instrumentation
framework \cite{valgrind}, within its VEX intermediate representation layer \cite{valgrind-tutorial}. It collects data
on five address-keyed profile families: conditional branch directions, conditional
branch outcomes by four-bit global-history context, dynamic branch targets, callgraph
edges, and CFG edges. All comparisons hold the binary fixed. Thus, addresses
identify the same instruction stream in both runs, and differences are
attributable to the execution inputs and their dynamic behavior.

We evaluate the methodology on SPEC CPU2017 and CPU2026 single-threaded
workloads \cite{cpu2017, cpu2026}.  The corpus includes 49 rate
train/reference pairs and seven refrate/refspeed pairs with the same rate
binary.  It contains high-fidelity controls where proxy and target behave
identically, as well as complex workloads where BBV support, branch targets,
call edges, or CFG edge mass diverge.  The case studies in
\autoref{sec:case-studies} demonstrate how our multi-view measurements
systematically diagnose these disparities across orthogonal execution
dimensions.

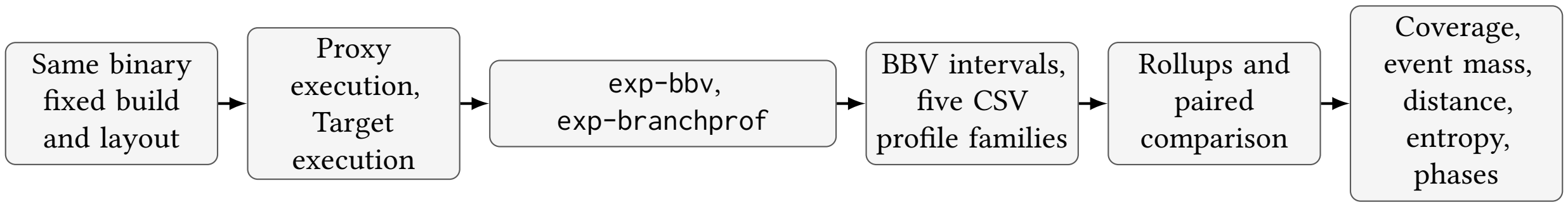
\begin{figure*}[t]
\centering
\begin{tikzpicture}[
  stage/.style={draw=black!65, rounded corners=3pt, align=center,
                minimum height=9mm, text width=20mm, fill=black!4},
  arrow/.style={->, thick, >=latex}
]
\node[stage] (binary) {Same binary\\fixed build and layout};
\node[stage, right=3mm of binary] (inputs) {Proxy execution,\\Target execution};
\node[stage, text width=34mm, right=3mm of inputs] (collect) {\texttt{exp-bbv},\\\texttt{exp-branchprof}};
\node[stage, right=3mm of collect] (profiles) {BBV intervals,\\five CSV profile families};
\node[stage, right=3mm of profiles] (rollup) {Rollups and\\paired comparison};
\node[stage, right=3mm of rollup] (report) {Coverage, event mass,\\distance, entropy, phases};
\draw[arrow] (inputs) -- (collect);
\draw[arrow] (binary) -- (inputs);
\draw[arrow] (collect) -- (profiles);
\draw[arrow] (profiles) -- (rollup);
\draw[arrow] (rollup) -- (report);
\end{tikzpicture}
\caption{Paired-execution workflow.  The comparison is meaningful
only when both inputs run the same binary under a controlled build and layout.
The result is a multi-view representativeness report, not a prediction of a
specific hardware or compiler outcome.}
\label{fig:workflow}
\Description{Workflow from a fixed binary and two inputs through Valgrind
collection, profile rollups, and a multi-metric representativeness report.}
\end{figure*}
\begin{table*}[t]
\footnotesize
\centering
\setlength{\aboverulesep}{0pt}
\setlength{\belowrulesep}{0pt}
\setlength{\extrarowheight}{.75ex}

\begin{tabularx}{\dimexpr\textwidth - 8pt\relax}{@{}
  >{\hsize=0.50\hsize}L
  >{\hsize=1.10\hsize}L
  >{\hsize=1.40\hsize}L
  @{}}
\arrayrulecolor{lightgray!40}
\rowcolor{tablehead}
\thead[l]{\textbf{\textcolor{white}{Family}}} &
\thead[l]{\textbf{\textcolor{white}{Key and Dynamic Weight}}} &
\thead[l]{\textbf{\textcolor{white}{Comparison Use}}} \\
\bottomrule
BBV & basic-block PC; interval and whole-run execution count &
phase recurrence, block support, weighted coverage, distribution overlap,
cosine similarity, and hot-block coverage \\
\bottomrule
Conditional branch & branch PC; taken and not-taken counts &
branch support, outcome bias, event coverage, and taken-probability error \\
\bottomrule
Branch history & branch PC and 4-bit history; taken and not-taken counts &
outcome structure conditioned on short history and entropy reduction \\
\bottomrule
Branch target & branch PC and target PC; transfer count &
target support, reference event-mass coverage, target distribution, and target
entropy error \\
\bottomrule
Callgraph & callsite PC and callee PC; call count &
dynamic call support, target fan-out, callsite hotness, and callee distribution \\
\bottomrule
CFG & source PC and destination PC; transfer count &
dynamic control-transfer support, successor distribution, and edge hotness \\
\bottomrule
\end{tabularx}
\\[+8pt]
\caption{Address-keyed observations and the primary questions they answer.
All dynamic counts are aggregated within a process; history state is maintained
per Valgrind thread.}
\label{tab:profile-families}
\par
\setlength{\aboverulesep}{0.6ex}
\setlength{\belowrulesep}{0.9ex}
\setlength{\extrarowheight}{0pt}
\end{table*}

This paper presents five contributions:
\begin{itemize}[leftmargin=11pt]
\item A measurement contract for comparing two executions of the
same binary using dynamic support, event mass, distributional agreement, and
phase behavior.
\item \texttt{exp-branchprof}, a Valgrind/VEX tool that collects five
address-keyed control-flow profile families from unmodified binaries.
\item Exact-oracle microbenchmarks and a reproducible validation harness for
conditional exits, inverted exits, calls, returns, direct and indirect jumps,
and multiprocess profile-file separation.
\item Rollups and comparisons for BBV, branch, history, target, callgraph, and
CFG data, together with a SPEC corpus that demonstrates representative and
non-representative execution pairs.
\item A privacy-preserving artifact workflow that returns aggregate summaries
for proprietary workload/proxy studies instead of raw PC-level profiles.
\end{itemize}

\section{Design Goals and Measurement Contract}
\label{sec:contract}

Our objective is to compare a cheaper execution \(T\) against a target
execution \(R\) of the \emph{same binary}.  The binary, ISA, compiler, linker
layout, relevant environment, and intended process scope must be controlled.
Controlling these execution factors is essential because address-keyed observations
can be joined across executions only when an address denotes the same instruction
stream.  The method is therefore appropriate for proxy/full-workload, train/reference,
or configuration/input comparisons, but not for comparing independently built
binaries without additional symbolization or normalization.

Each profile family is a weighted event multiset.  As summarized in
\autoref{tab:profile-families}, a key \(k\) identifies a basic block,
branch, branch/history pair, branch/target pair, call edge, or CFG edge,
and \(c_X(k)\) is its dynamic count in execution \(X\).  The workflow in
\autoref{fig:workflow} evaluates which keys occur, how much target event mass
the proxy covers, whether shared keys have similar relative weights, and whether
the temporal BBV phase structure recurs.

The method does \emph{not} directly model cache behavior, hardware predictor
aliasing, speculative execution, compilation profitability, or source-level
semantics.  It also does not declare a proxy representative from one threshold.
Instead, it exposes a compact evidence set that a downstream user can interpret
for the decision at hand.  High support coverage with low event-mass coverage,
for example, indicates that a proxy reaches many target identities but misses
where the target spends time.

\section{BranchProf Design and Implementation}
\label{sec:implementation}
We describe the design and output structures of \texttt{exp-branchprof},
including Valgrind-specific details.

\subsection{Tool Architecture}

\texttt{exp-branchprof} is an experimental Valgrind tool implemented at the
VEX IR level. It instruments an unmodified
client binary and records dynamic control transfers as address-keyed counts.
This choice permits the same executable to be used for native timing, BBV
collection with \texttt{exp-bbv} \cite{weaver-hipeac2008, valgrind-expbbv, valgrind}, and
BranchProf collection. It also avoids source-language, debug-information, and
build-system dependencies. As with other address-based dynamic tools, an
address is meaningful across two runs only when both executions use the same
binary image and layout.

The tool maintains five OSetGen tables and writes one CSV file per table at
process termination. They respectively contain conditional branch direction
counts (branch PC plus taken and not-taken columns), history-conditioned
direction counts (branch PC and history), dynamic branch-target counts, dynamic
call edges, and dynamic CFG edges. The last three use source/destination PC
pairs and a count; callgraph source and destination are named caller and callee.
The output filenames accept Valgrind's
\texttt{\%p} expansion and therefore should include it whenever the client may
fork. The artifact documents the command-line options and schemas.

\subsection{Conditional Exits and Short History}

VEX represents a conditional control transfer with an \texttt{Ist\_Exit}
statement. For each exit, BranchProf inserts one helper on the guarded path and
one on the complementary fall-through path, keyed by the control-transfer
instruction address. Some lifted conditional exits have an exit destination
equal to the architectural next instruction. In that representation, the
guarded VEX path corresponds to the architectural fall-through. BranchProf
detects this inverted-exit case and swaps taken/not-taken accounting, following
the convention used by Valgrind's \texttt{lackey} accounting.

The history table adds a four-bit global branch-history value to the same
branch-PC key. History is maintained per Valgrind thread, avoiding a shared
history state across threads. It provides a compact descriptive context rather
than a hardware branch-predictor simulation, omitting predictor-table aliasing,
speculative updates, and misprediction penalties.

\subsection{Targets and Dynamic Edges}

For each VEX conditional exit, BranchProf emits a CFG edge from the instruction
address to the observed destination. It also instruments the final exit of a
translated block when the VEX jump kind is \texttt{Ijk\_Boring},
\texttt{Ijk\_Call}, or \texttt{Ijk\_Ret}. Here, ``boring'' is VEX terminology
for an ordinary non-call, non-return control transfer; it includes direct and
indirect jumps. This handling records ordinary fall-throughs and jumps, calls,
returns, and indirect destinations in the CFG profile.

\begin{figure}[t]
  \centering
  \includegraphics[width=\columnwidth]{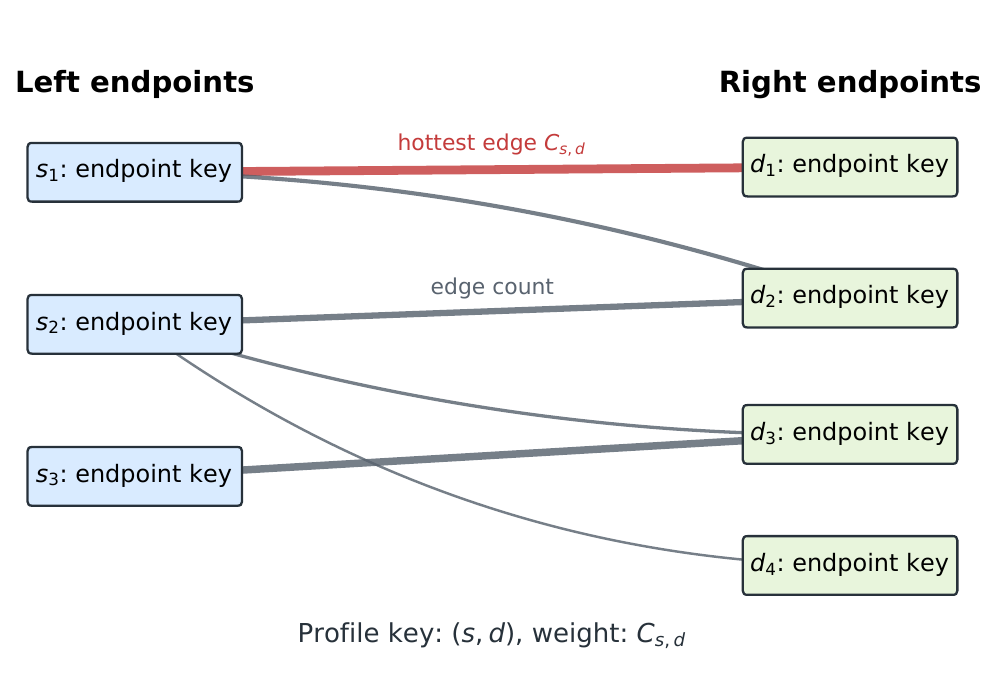}
  \caption{Dynamic edge representation used by BranchProf. For callgraph
  profiles, \(s\) is a callsite and \(d\) is a callee; for CFG profiles, they
  are a transfer source and observed destination. Edge thickness represents
  dynamic count.}
  \label{fig:dynamic-edge-schematic}
  \Description{Weighted dynamic edges for callgraph and CFG profiles.}
\end{figure}

For indirect final exits, the runtime destination expression is supplied to the
helper, producing a dynamic branch-target distribution. Direct constant
destinations are passed directly. Call exits additionally produce a callgraph
edge from the callsite to the dynamic callee. Returns are CFG and branch-target
events but are not callgraph calls. The common edge model is shown in
\autoref{fig:dynamic-edge-schematic}. This lighter-weight,
comparison-oriented representation follows the feasibility demonstrated by
Valgrind-based CFG reconstruction, while not attempting full
per-function CFG recovery \cite{cfggrind}.

\subsection{Process and Thread Scope}

Profile tables are process-local and files are emitted when each process exits.
The recommended \texttt{\%p} filenames prevent a child from overwriting its
parent's output. A later rollup may combine process files only when that matches
the workload question. Within one process, the event tables aggregate all
threads, while branch history remains per thread. This makes the tool usable for
multithreaded clients, but the resulting aggregate reflects all threads and can
be affected by Valgrind-induced schedule changes. We therefore evaluate only
single-threaded SPEC runs and do not make thread-level representativeness or
repeatability claims.

\subsection{BBV Collection}

BBVs complement BranchProf rather than duplicate it. \texttt{exp-bbv} collects
sparse basic-block vectors over fixed instruction intervals. We use one billion
instructions per interval in the SPEC study, concatenate the proxy and target
interval sequences, and construct an all-pairs self-similarity matrix.
Whole-run BBV histograms provide an address-keyed block distribution that can be
compared by the same weighted-event machinery as the BranchProf families. The
artifact's small address-output adjustment for \texttt{exp-bbv} allows the
histogram keys to be joined across executions of the fixed binary.

\section{Comparison Methodology}
\label{sec:comparison}

Each profile family is a weighted multiset of dynamic event keys. Let
\(c_X(k)\) be the dynamic count for key \(k\) in execution \(X\), and let
\(K_X=\{k:c_X(k)>0\}\). We compare a proxy \(T\) against a target \(R\). This
orientation is important because coverage asks what target behavior the proxy
observes, whereas accuracy asks how much proxy behavior occurs in the target.

\subsection{Support, Event Mass, and Distribution}

The unweighted support measures are
\begin{equation}
  \mathrm{coverage}=\frac{|K_T\cap K_R|}{|K_R|},\qquad
  \mathrm{accuracy}=\frac{|K_T\cap K_R|}{|K_T|}.
\end{equation}
As illustrated in \autoref{fig:coverage-accuracy-venn}, both metrics share the
same set of overlapping keys but use different reference denominators. Coverage
evaluates what fraction of target keys the proxy exercises, whereas accuracy
measures how much of the proxy's observed execution belongs to the target.
These unweighted set intersections distinguish an execution that reaches many
target identities from one that observes unrelated identities, but neither
indicates how frequently those identities execute. Let \(C_X=\sum_k c_X(k)\).
We therefore report reference event-mass coverage and proxy event-mass accuracy:

\begin{figure}[t]
  \centering
  \includegraphics[width=0.9\columnwidth]{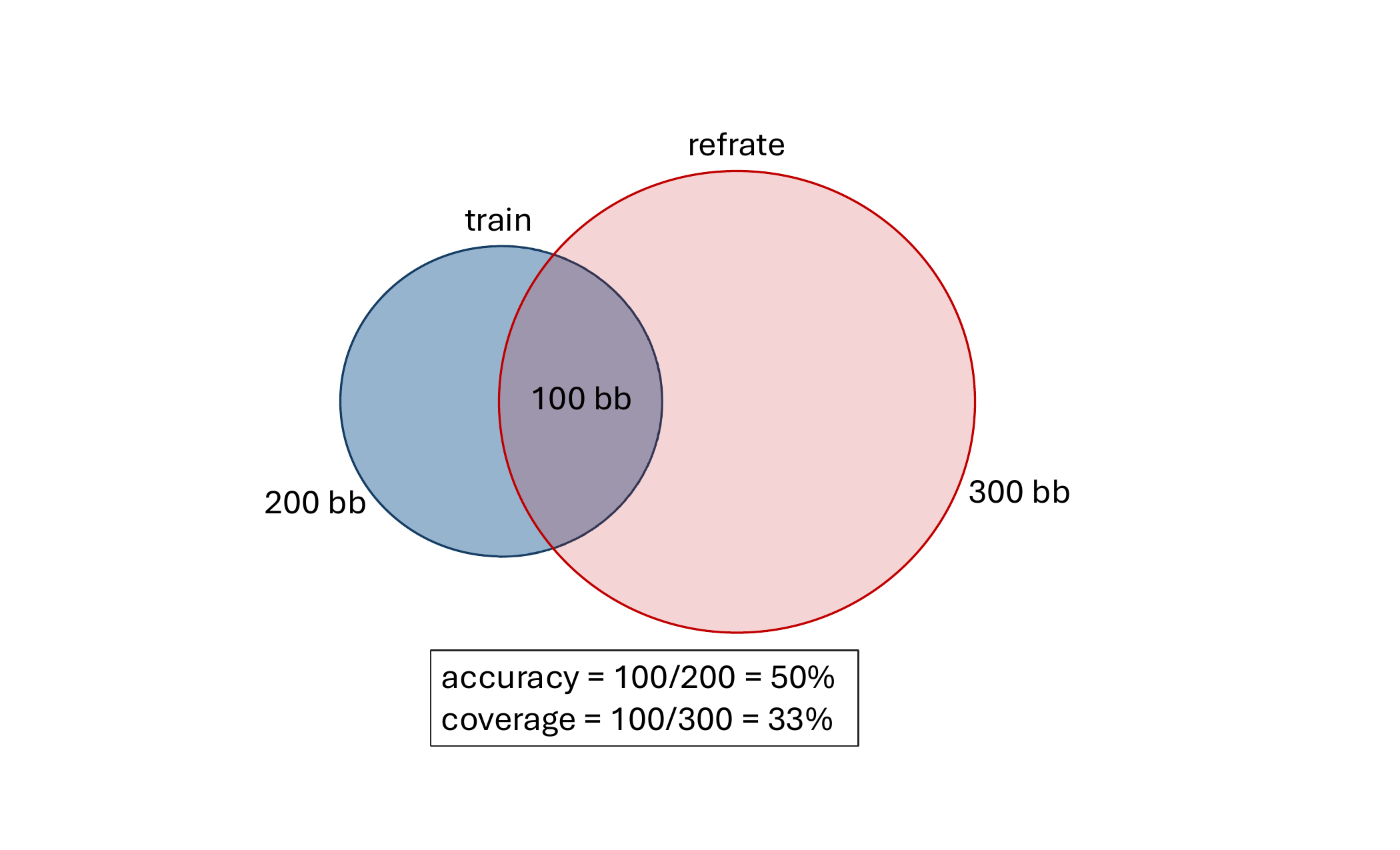}
  \caption{Support coverage and accuracy use the same shared-key set with
  different denominators. In this example, the train proxy has 200 basic-block
  keys, the refrate target has 300, and 100 keys are shared; accuracy is
  \(100/200=50\%\), while coverage is \(100/300=33\%\).}
  \Description{Venn diagram showing 200 train basic-block keys and 300 refrate
  basic-block keys with 100 shared keys. It illustrates accuracy as shared keys
  divided by train keys and coverage as shared keys divided by refrate keys.}
  \label{fig:coverage-accuracy-venn}
\end{figure}

\begin{equation}
  E_R=\frac{\sum_{k\in K_T\cap K_R}c_R(k)}{C_R},\qquad
  E_T=\frac{\sum_{k\in K_T\cap K_R}c_T(k)}{C_T}.
\end{equation}
\(E_R\) is the principal coverage statistic in this paper, representing the
fraction of target dynamic events carried by keys seen in the proxy.

Support can agree while relative weights disagree. With normalized distributions
\(p_X(k)=c_X(k)/C_X\), we use distribution overlap and total variation distance:
\begin{equation}
  \mathrm{overlap}=\sum_{k\in K_T\cup K_R}\min(p_T(k),p_R(k)),\qquad
  \mathrm{TVD}=1-\mathrm{overlap}.
\end{equation}
We also report cosine similarity for whole-run BBV histograms. These
probability-distribution views are related to prior information-theoretic
profile comparison work \cite{savari_combining_profiles_2000}, but we use them
as descriptive evidence rather than profile blending inputs.

\subsection{BBV Phase and Hot-Region Views}

For interval vector \(x_i\), we normalize to unit \(L_2\) norm and compare
intervals with
\begin{equation}
  D(i,j)=\left\lVert
  \frac{x_i}{\lVert x_i\rVert_2} -
  \frac{x_j}{\lVert x_j\rVert_2}\right\rVert_2.
\end{equation}
The resulting self-similarity matrix preserves execution order. Its
off-diagonal proxy/target regions show whether target phases recur in the
proxy, a use of BBVs complementary to SimPoint's phase-selection objective
\cite{bbv_distribution, simpoints, simpoints3}. We additionally report
weighted block coverage and hot-block coverage. Let \(H_R(0.90)\) be the
smallest set of target blocks, ordered by decreasing \(c_R(b)\), whose
cumulative count reaches \(90\%\) of target BBV weight. We report both its
support coverage and the target mass carried by shared hot blocks:
\begin{equation}
  \frac{|H_R(0.90)\cap K_T|}{|H_R(0.90)|},\qquad
  \frac{\sum_{b\in H_R(0.90)\cap K_T}c_R(b)}
       {\sum_{b\in H_R(0.90)}c_R(b)}.
\end{equation}
This catches the common case where most blocks are shared but the proxy omits
the target's hot regions.

\subsection{Outcome and Fan-Out Entropy}

For a conditional branch with taken probability \(p_b\), binary entropy is
\begin{equation}
  H(p_b)=-p_b\log_2 p_b-(1-p_b)\log_2(1-p_b).
\end{equation}
For each four-bit history \(h\), let \(C_{b,h}\) be the dynamic count and
\(p_{b,h}\) the taken probability of branch \(b\). The history-conditioned
entropy and the reduction attributed to this short context are
\begin{equation}
  H_{\mathrm{hist}}(b)=\sum_h\frac{C_{b,h}}{C_b}H(p_{b,h}),\qquad
  \Delta H(b)=H(p_b)-H_{\mathrm{hist}}(b),
\end{equation}
where \(C_b=\sum_h C_{b,h}\). The rollups report errors in \(\Delta H\)
weighted by reference event counts, not a branch-predictor simulation.

Callgraph and CFG profiles use a common weighted dynamic-edge representation
\((s,d,C_{s,d})\), where \(s\) and \(d\) denote a callsite and callee for
callgraph profiles, or a transfer source and observed destination for CFG
profiles. Their event volume is \(C=\sum_{s,d}C_{s,d}\). We report the dynamic
edge surface as unique edges
\(E=|\{(s,d):C_{s,d}>0\}|\), source endpoints
\(S=|\{s:\exists d, C_{s,d}>0\}|\), and destination endpoints
\(D=|\{d:\exists s, C_{s,d}>0\}|\). These quantities describe the dynamic
graph exercised by an input, not the complete static program graph.

For a branch target, callsite target, or CFG successor distribution with
dynamic probabilities \(p_{s,d}\), fan-out entropy is
\begin{equation}
  H(s)=-\sum_d p_{s,d}\log_2 p_{s,d}.
\end{equation}
All entropy values are in bits. A zero value means that the observed source was
monomorphic; greater values indicate more balanced observed fan-out. Rollups
report errors in probabilities and entropy, weighted by reference event counts,
over keys common to both runs. These quantities explain \emph{how} shared control-flow
sites differ, while coverage and TVD establish whether the relevant support and
dynamic mass are present.

\subsection{Interpretation}

No metric alone certifies representativeness. High support coverage with low
\(E_R\) means that the proxy reaches target code but misses where the target
spends time. High \(E_R\) with high TVD means the proxy covers target events
but weights them differently. High branch-target coverage and low callgraph
coverage can identify a call-structure mismatch that a target-only view would
miss. We therefore interpret the profile families jointly, with downstream
experiments used to test decision-specific usefulness rather than to redefine
the measurement contract.

\section{Validation and Operational Characteristics}
\label{sec:validation}

\subsection{Exact Semantic Oracles}

We validate BranchProf with deterministic, non-PIE microbenchmarks on AArch64 and x86-64,
whose assembly labels identify the expected transfer PCs. The harness executes
the native and instrumented program, checks identical stdout, resolves labels
with \texttt{nm}, and validates exact CSV rows. It covers VEX-level cases
that can be overlooked in source-level testing, including inverted conditional
exits, final exits with ordinary \texttt{Ijk\_Boring} jumps, direct and indirect
targets, return edges, forked processes, and multithreaded aggregation with
per-thread history. The checks use exact expected counts, not sampling estimates;
\autoref{tab:semantic-validation} gives the complete expected-versus-observed
matrix. The accompanying artifact packages the tests and checkers.

The accompanying broader microbenchmark exercises all five profile families
with nontrivial counts and a deterministic indirect-call distribution. It is a
smoke test for the full collection and output pipeline; the exact-oracle suite
is the semantic test. Both suites and their checkers are packaged with the
artifact.

\subsection{Evaluation Platform}

All SPEC experiments use the platform in \autoref{tab:evaluation-platform}.
The native, BBV, and BranchProf executions use the same fixed AArch64 binary;
the reported SPEC rate results are single-threaded. This controls the
address-key join used by the comparison and makes the operational measurements
directly comparable.

\begin{table}[t]
\footnotesize
\centering
\caption{Evaluation platform and build configuration.}
\label{tab:evaluation-platform}
\setlength{\aboverulesep}{0pt}
\setlength{\belowrulesep}{0pt}
\setlength{\extrarowheight}{.75ex}
\begin{tabularx}{\dimexpr\columnwidth - 8pt\relax}{@{}
  >{\hsize=0.8\hsize}L
  >{\hsize=1.20\hsize}L
  @{}}
\arrayrulecolor{lightgray!40}
\rowcolor{tablehead}
\thead[l]{\textbf{\textcolor{white}{Component}}} &
\thead[l]{\textbf{\textcolor{white}{Configuration}}} \\
\bottomrule
Processor & AmpereOne A192-32X \cite{AmpereOne_Brief, ampereone_perf} \\
\bottomrule
Cores & 192 Ampere cores (AArch64, ARMv8.6) \\
\bottomrule
Frequency & 3.2 GHz \\
\bottomrule
Memory & 512 GB DDR5-5600 \\
\bottomrule
%L1 Cache & 16 KiB instruction + 64 KiB data \\
%\bottomrule
%L2 Cache & 2 MiB \\
%\bottomrule
%L3 Cache & 64 MiB SLC \\
%\bottomrule
Compiler and Flags & GCC 16.1 \texttt{-O3} \\
\bottomrule
Operating Environment & Ubuntu 24.04.4 LTS \\ & Linux kernel 6.8.0-101-generic \\
\bottomrule
NUMA & 1 node per socket (monolithic) \\
\bottomrule
\end{tabularx}
\par
\setlength{\aboverulesep}{0.6ex}
\setlength{\belowrulesep}{0.9ex}
\setlength{\extrarowheight}{0pt}
\end{table}

\subsection{Runtime Cost and Scaling}

BranchProf is intended for offline workload characterization, not online
production tracing. \autoref{tab:overhead} summarizes the matched SPEC runs.
Across the two suites' train and rate-reference collections, the median slowdown
ranges from \(45.4\times\) to \(48.5\times\); the CPU2017 refrate median is
\(46.6\times\), and CPU2026 refrate is \(45.6\times\). The refspeed subsets
are smaller and can be more expensive, so they are not used as a suite-wide
cost claim.

\begin{table}[t]
\footnotesize
\centering
\setlength{\aboverulesep}{0pt}
\setlength{\belowrulesep}{0pt}
\setlength{\extrarowheight}{.75ex}
\caption{BranchProf slowdown in matched native and instrumented SPEC runs. The CPU2026 harness also reports RSS memory sizes in its log files.}
\label{tab:overhead}
\begin{tabularx}{\dimexpr\columnwidth - 8pt\relax}{@{}
  >{\hsize=0.75\hsize}L
  >{\hsize=0.60\hsize}L
  >{\hsize=0.20\hsize}C
  >{\hsize=1.72\hsize}C
  >{\hsize=1.72\hsize}C
  @{}}
\arrayrulecolor{lightgray!40}
\rowcolor{tablehead}
\thead[l]{\textbf{\textcolor{white}{Suite}}}&
\thead[l]{\textbf{\textcolor{white}{Input}}}&
\thead[c]{\textbf{\textcolor{white}{\(n\)}}}&
\thead[c]{\textbf{\textcolor{white}{Perf Slowdown}}}&
\thead[c]{\textbf{\textcolor{white}{Peak Mem (RSS)}}}\\
 & & & \textbf{geomean\(/\)median} & \textbf{geomean\(/\)median} \\
\bottomrule
CPU2017 & train & 33 & \(45.0\times/~48.5\times\) & -- \\
\bottomrule
CPU2017 & refrate & 23 & \(35.9\times/~46.6\times\) & -- \\
\bottomrule
CPU2026 & train & 31 & \(53.8\times/~45.4\times\) & \(1.75\times/~1.48\times\) \\
\bottomrule
CPU2026 & refrate & 26 & \(45.9\times/~45.6\times\) & \(1.33\times/~1.07\times\) \\
\bottomrule
\end{tabularx}
\par
\setlength{\aboverulesep}{0.6ex}
\setlength{\belowrulesep}{0.9ex}
\setlength{\extrarowheight}{0pt}
\end{table}

The 128 matched BranchProf records show that dynamic event rate, rather than
static profile size alone, best explains the overhead. CFG transfers per native
second have Spearman correlation \(0.948\) with slowdown; calls per native
second have correlation \(0.914\). The complete corpus contains 142
input-profile sets, or 710 raw BranchProf CSV files, totaling \(1.41\) GB.
Branch-history and CFG-edge profiles dominate storage at \(458\) MB and
\(447\) MB, respectively; their median file sizes are \(0.95\) MB and
\(0.90\) MB, while their 90th-percentile sizes are \(9.63\) MB and
\(9.47\) MB. These are practical artifact-scale figures, not a universal
per-run storage estimate.

For the 52 CPU2026 rate pairs with harness-recorded peak RSS, BranchProf's
geomean/median RSS ratio is \(1.75\times/1.48\times\) on train and
\(1.33\times/1.07\times\) on refrate. The corresponding median increases are
\(53.1\) MiB and \(54.8\) MiB. The largest ratio is \(10.8\times\) for the refrate size of
\texttt{748.flightdm}, where the native peak is only \(4.5\) MiB;
the absolute increase is \(44.1\) MiB. These measurements are peak values
reported by the SPEC harness for the complete native or Valgrind-launched
execution, not the size of the emitted profile files.

\subsection{Repeated-Run Stability}

We repeated both train and refrate profile collection three times for three
CPU2026 workloads with large train/reference fidelity gaps: 729.abc, 735.gem5,
and 777.zstd.  This produces three within-input pairs for each combination of \{benchmark, input, profile-family\} and nine train/refrate pairs.  Across
the five BranchProf families, the 90 within-input comparisons have event
coverage of at least \(0.9999999827\) and TVD at most \(0.000159\).  The
largest variation is the 735.gem5 refrate branch-history profile; but its total
event-count difference is only \(0.01124\%\). Across the branch and
branch-history profiles, the largest mean absolute difference (MAE) in taken
probability between repeated runs, weighted by event counts from the second
run, is \(2.02 \times 10^{-5}\).

These bounds do not overlap with the input effect.  Every same-input repeat
pair has TVD at most \(0.000159\), whereas every train/refrate pair has TVD at
least \(0.391\), a separation of more than \(2{,}400\times\).  Similarly, the
worst repeated-run taken-probability MAE is more than \(2{,}200\times\) smaller
than the smallest cross-input MAE.  Same-input event coverage remains above
\(0.99999998\), while cross-input coverage never exceeds \(0.748\).  Thus,
changing the input produces a clear, reproducible profile difference that is
far larger than collection variation for these workloads.  This is a targeted
single-threaded replication study, rather than a claim that every workload or
multithreaded execution has the same determinism.

\subsection{Scope and Limitations}

The fixed ISA and layout make the address joins well defined; they do not limit
the method to AArch64, as we have also validated the entire flow on x86-64. The tool does not infer source-level names, static CFG
completeness, cache behavior, predictor mispredictions, or compiler
profitability. Cachegrind's branch simulation, for example, evaluates a
particular simple predictor model \cite{cachegrind_manual}; BranchProf instead
records raw dynamic event distributions for paired-execution comparison.

The tables aggregate all client threads in a process and retain a separate
four-bit history per Valgrind thread. Multithreaded execution is therefore
collectible, but scheduler perturbation and interthread aggregation complicate
interpretation. Users should choose process scope deliberately, use \texttt{\%p}
filenames for multiprocess clients, and avoid comparing different binary builds
without an additional address-normalization layer.

\section{Case Studies in Workload Representativeness}
\label{sec:case-studies}

We apply the workflow to 49 SPEC CPU2017 and CPU2026 rate train/refrate pairs
and seven refrate/refspeed pairs \cite{cpu2017, cpu2026}. The goal is not to
rank every benchmark with one score. Rather, the cases demonstrate how our
multi-view report distinguishes close pairs, broad dynamic mismatches, and
family-specific counterexamples. \autoref{tab:case-studies} summarizes six
archetypal pairings, each directly illustrated by an empirical phase recurrence
visualization in the paper. It reports BBV coverage alongside all five
BranchProf families; cells are shaded with a gradient reflecting their deficit
from complete coverage (\(1.000\), white). The full per-benchmark rollups and
comprehensive data tables are provided in the accompanying artifact.

\ifdefined\gradient\else\fi
\begin{table*}[t]
\footnotesize
\centering
\setlength{\aboverulesep}{0pt}
\setlength{\belowrulesep}{0pt}
\setlength{\extrarowheight}{.75ex}
\caption{Selected paired-execution coverage reports. \(E\) is target
event-mass coverage by proxy-observed keys. The five BranchProf columns are
conditional-branch direction (Br.), branch history (Hist.), branch target
(Tgt.), callgraph (CG), and CFG edge (CFG); BBV values are weighted coverage.}
\label{tab:case-studies}
\begin{tabularx}{\dimexpr\textwidth - 8pt\relax}{@{}
  >{\hsize=0.85\hsize}L
  >{\hsize=0.90\hsize}L
  >{\hsize=0.55\hsize}C
  >{\hsize=0.55\hsize}C
  >{\hsize=0.55\hsize}C
  >{\hsize=0.55\hsize}C
  >{\hsize=0.55\hsize}C
  >{\hsize=0.55\hsize}C
  >{\hsize=1.80\hsize}L
  @{}}
\arrayrulecolor{lightgray!40}
\rowcolor{tablehead}
\thead[l]{\textbf{\textcolor{white}{Benchmark}}} &
\thead[l]{\textbf{\textcolor{white}{Comparison}}} &
\thead[c]{\textbf{\textcolor{white}{BBV}}} &
\thead[c]{\textbf{\textcolor{white}{Br. \(E\)}}} &
\thead[c]{\textbf{\textcolor{white}{Hist. \(E\)}}} &
\thead[c]{\textbf{\textcolor{white}{Tgt. \(E\)}}} &
\thead[c]{\textbf{\textcolor{white}{CG \(E\)}}} &
\thead[c]{\textbf{\textcolor{white}{CFG \(E\)}}} &
\thead[l]{\textbf{\textcolor{white}{Interpretation}}} \\
\bottomrule
\texttt{709.cactus\_r} & train / refrate & \gradient{1.000} & \gradient{1.000} & \gradient{1.000} & \gradient{1.000} & \gradient{1.000} & \gradient{1.000} &
high-fidelity control \\
\bottomrule
\texttt{782.lbm\_r} & train / refrate & \gradient{1.000} & \gradient{1.000} & \gradient{1.000} & \gradient{1.000} & \gradient{1.000} & \gradient{1.000} &
high-fidelity control \\
\bottomrule
\texttt{729.abc\_r} & train / refrate & \gradient{0.488} & \gradient{0.482} & \gradient{0.382} & \gradient{0.735} & \gradient{0.673} & \gradient{0.488} &
broad event-mass mismatch \\
\bottomrule
\texttt{735.gem5\_r} & train / refrate & \gradient{0.547} & \gradient{0.538} & \gradient{0.521} & \gradient{0.567} & \gradient{0.510} & \gradient{0.538} &
target and edge mismatch \\
\bottomrule
\texttt{519.lbm\_r} & train / refrate & \gradient{0.990} & \gradient{1.000} & \gradient{0.985} & \gradient{0.999} & \gradient{0.515} & \gradient{0.992} &
callgraph threshold failure \\
\bottomrule
\texttt{729.abc\_r} & refrate / refspeed & \gradient{1.000} & \gradient{1.000} & \gradient{1.000} & \gradient{0.999} & \gradient{0.998} & \gradient{0.999} &
high coverage, high BBV TVD \\
\bottomrule
\end{tabularx}
\par
\setlength{\aboverulesep}{0.6ex}
\setlength{\belowrulesep}{0.9ex}
\setlength{\extrarowheight}{0pt}
\end{table*}

\subsection{Controls and Mismatches}

\begin{figure}[t]
  \centering
  \begin{subfigure}[b]{\columnwidth}
    \centering
    \includegraphics[width=\linewidth]{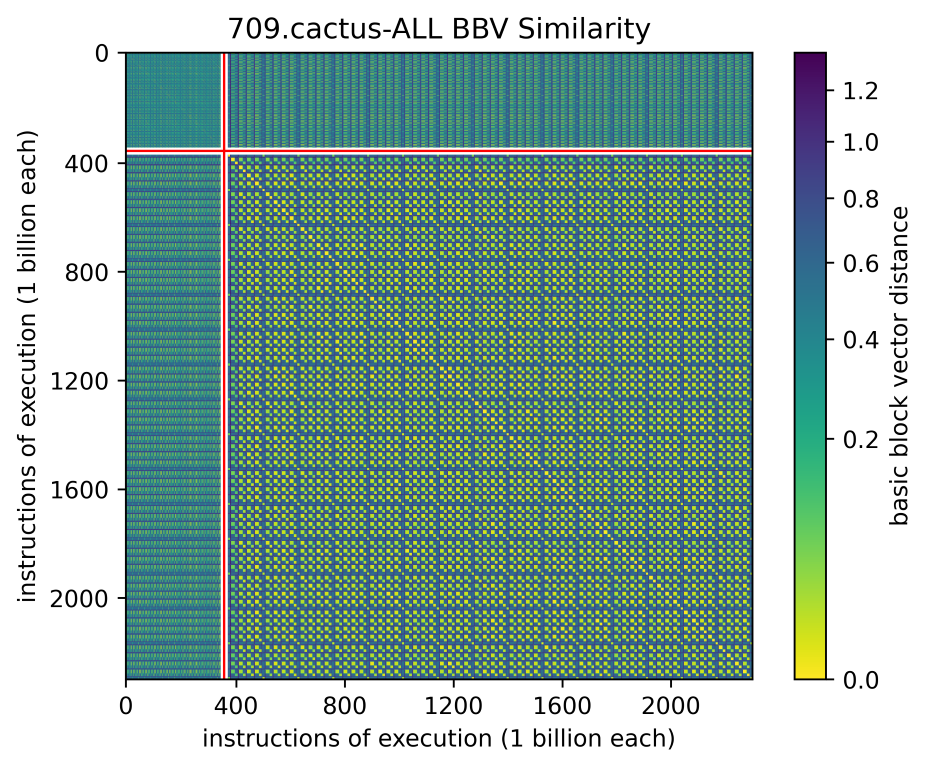}
    \caption{\texttt{709.cactus} (High-Fidelity Control)}
    \label{fig:cactus-bbv}
  \end{subfigure}\\[6pt]
  \begin{subfigure}[b]{\columnwidth}
    \centering
    \includegraphics[width=\linewidth]{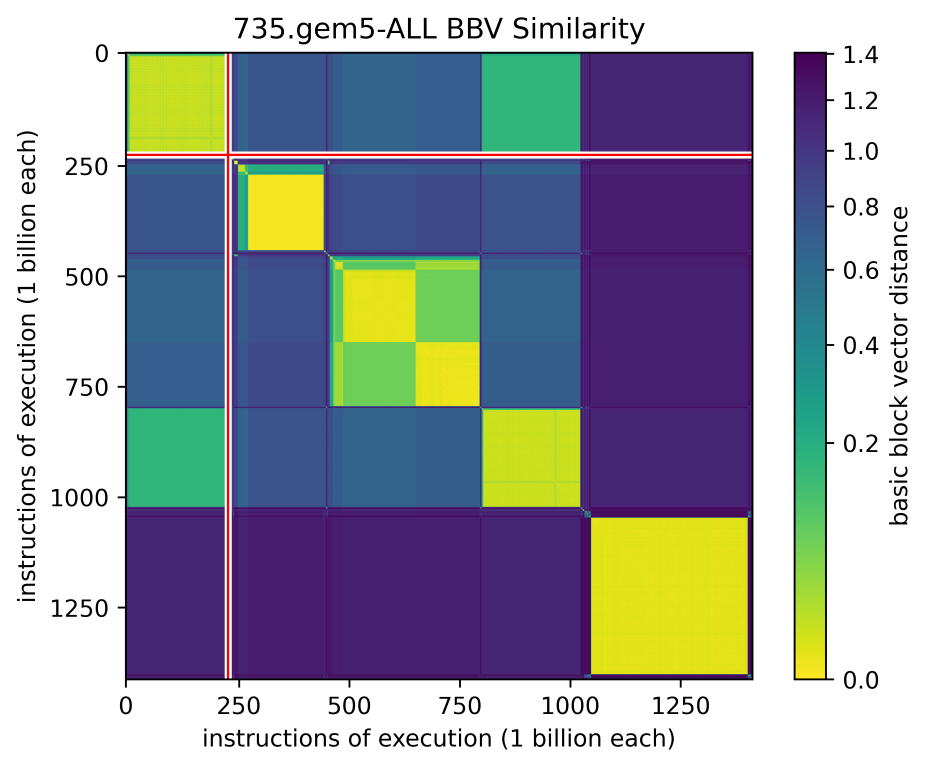}
    \caption{\texttt{735.gem5} (Structural Mismatch)}
    \label{fig:gem5-bbv}
  \end{subfigure}
  \caption{BBV phase self-similarity contrasting a high-fidelity control against a structural mismatch (train proxy concatenated with reference target). In (a) \texttt{709.cactus}, all dynamic phases recur ($100\%$ weighted coverage). In (b) \texttt{735.gem5}, prominent reference phases (dark off-diagonal regions) are entirely unexercised by the proxy ($54.7\%$ coverage).}
  \label{fig:control-vs-mismatch}
  \Description{Vertical comparison of BBV self-similarity matrices for 709.cactus and 735.gem5.}
\end{figure}

The CPU2026 \texttt{709.cactus} and \texttt{782.lbm} pairs in
\autoref{tab:case-studies} (rows 1 and 2) represent high-fidelity controls. Their whole-run BBV
weighted coverage and hot-block coverage are \(100\%\), and all reported
five BranchProf event-mass coverages are \(1.000\). As shown in \autoref{fig:cactus-bbv}
and \autoref{fig:782-lbm-bbv}, the off-diagonal recurrence mirrors the
diagonal, showing that every dynamic phase observed during reference execution recurs
during the train proxy. This unanimous agreement across independently collected BBV and
control-flow views is the empirical signature expected of a useful proxy.

In contrast, \texttt{729.abc} and \texttt{735.gem5} are broad mismatches, and demonstrate that
nominal code reachability can mask dynamic event-mass collapse. For
\texttt{729.abc}, the train run covers only \(48.8\%\) of dynamic reference
BBV mass; its BranchProf coverage ranges from \(0.382\) for branch history to
\(0.735\) for branch targets (\autoref{tab:case-studies}, row 3). As illustrated in the
front-page recurrence plot (\autoref{fig:abc-bbv}, Page~\pageref{fig:abc-bbv}),
the training proxy exercises an isolated subset of logic-synthesis commands
\cite{729-abc, 729-abc-doc}, leaving over half of the reference workload's
execution mass unobserved.

Similarly, \texttt{735.gem5} reaches most of its proxy support but omits
substantial target event mass. Its BBV weighted coverage is \(54.7\%\), while
the five BranchProf families range from \(0.510\) callgraph coverage to \(0.567\)
branch-target coverage (\autoref{tab:case-studies}, row 4). The BBV phase view in
\autoref{fig:gem5-bbv} visualizes this
mismatch temporally. Several prominent target execution regions (the dark
off-diagonal blocks) lack a close proxy counterpart because the train input
exercises a restricted architectural component, leaving major simulation
subsystems inactive.

\subsection{Why Multiple Families Matter}

A single observation family—whether basic-block vectors, branch targets, or
static reachability—cannot reliably certify workload representativeness. The
remaining pairings in \autoref{tab:case-studies} illustrate why multi-family
profiling is necessary.

\begin{figure}[t]
  \centering
  \begin{subfigure}[b]{\columnwidth}
    \centering
    \includegraphics[width=\linewidth]{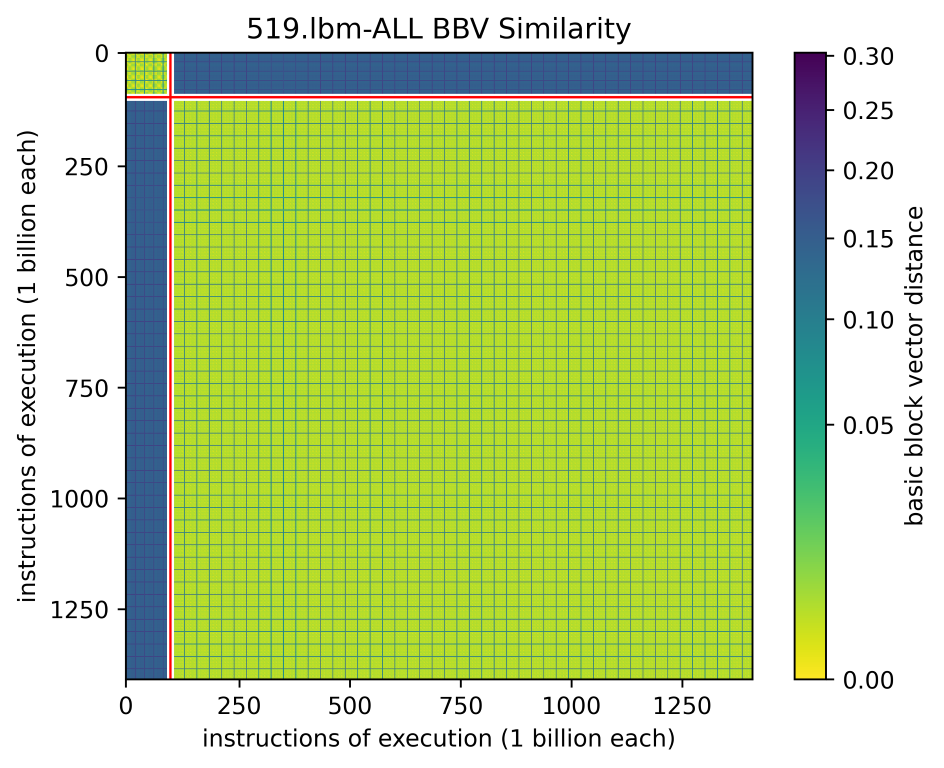}
    \caption{\texttt{519.lbm} (CPU2017: High BBV Match, Callgraph Divergence)}
    \label{fig:519-lbm-bbv}
  \end{subfigure}\\[6pt]
  \begin{subfigure}[b]{\columnwidth}
    \centering
    \includegraphics[width=\linewidth]{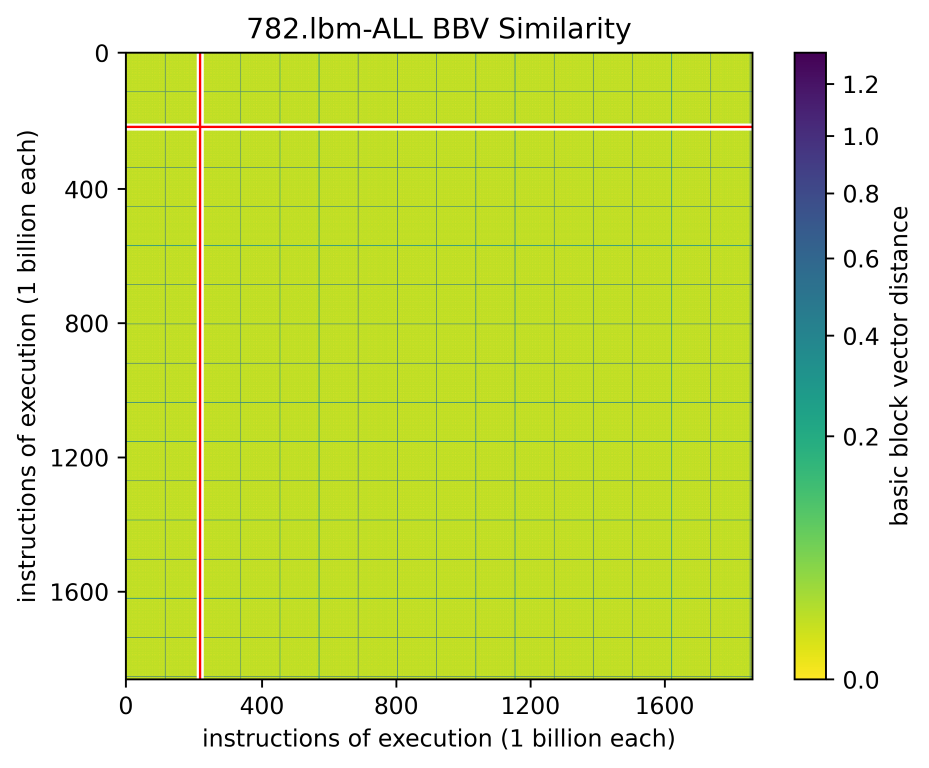}
    \caption{\texttt{782.lbm} (CPU2026: Uniform High Fidelity)}
    \label{fig:782-lbm-bbv}
  \end{subfigure}
  \caption{BBV phase self-similarity for \texttt{519.lbm} (CPU2017) and \texttt{782.lbm} (CPU2026). Both workloads exhibit near-perfect basic-block recurrence between train and reference ($99.0\%$ and $100\%$ BBV coverage). However, while \texttt{782.lbm} maintains complete fidelity across all BranchProf families ($1.000$ callgraph and CFG coverage), \texttt{519.lbm} conceals a major callgraph divergence ($0.515$ coverage), demonstrating why BBV recurrence alone cannot certify proxy fidelity without fine-grained control-flow profiling.}
  \label{fig:lbm-bbv-comparison}
  \Description{Vertical comparison of BBV self-similarity matrices for 519.lbm and 782.lbm.}
\end{figure}

CPU2017 \texttt{519.lbm} and CPU2026 \texttt{782.lbm} provide a directly
didactic contrast across two generations of the same fluid-dynamics application.
As shown in \autoref{fig:lbm-bbv-comparison}, both benchmarks exhibit
near-perfect basic-block phase recurrence between train and reference
(\autoref{fig:519-lbm-bbv} and \autoref{fig:782-lbm-bbv}), covering \(99.0\%\)
and \(100\%\) of dynamic BBV event mass, respectively. If evaluated by basic-block vectors alone, both would be certified
as high-fidelity proxies. However, while \texttt{782.lbm} achieves \(1.000\)
event coverage across all five BranchProf families (\autoref{tab:case-studies}, row 2),
\texttt{519.lbm} covers only \(0.515\) of dynamic reference callgraph event
mass (\autoref{tab:case-studies}, row 5), despite \(0.999\) branch-target
coverage and \(0.992\) CFG-edge coverage. A basic-block or branch-target view
alone misses this structural shift in dynamic call behavior. The profile families
are not competing similarity scores; they expose distinct, orthogonal ways a
proxy can be incomplete.

\begin{figure}[t]
  \centering
  \includegraphics[width=\columnwidth]{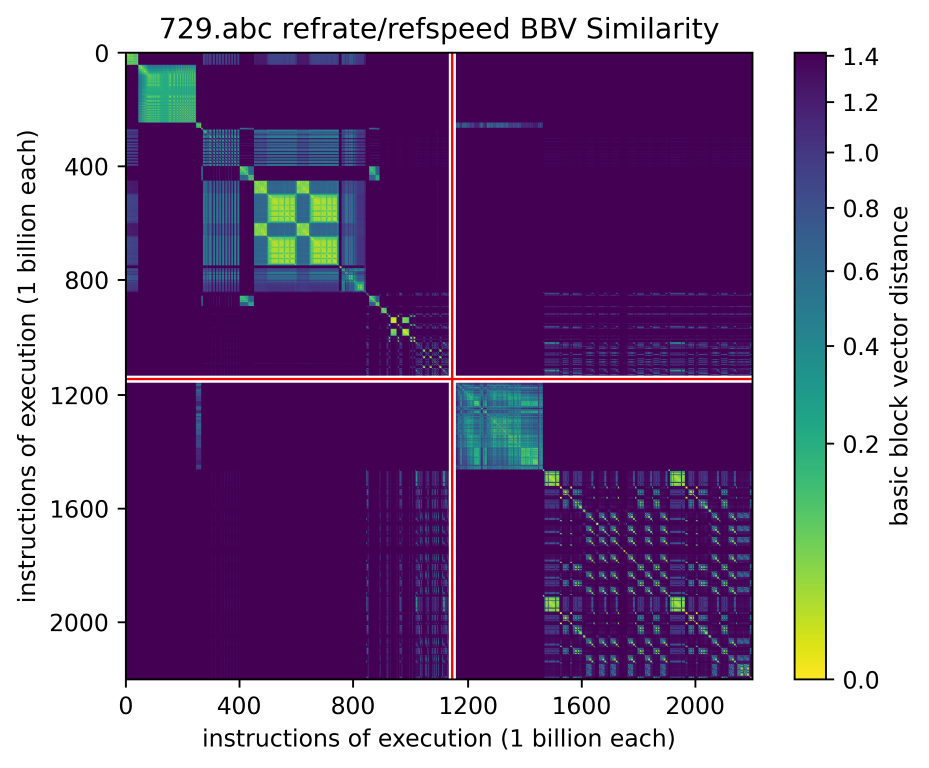}
  \caption{BBV phase self-similarity for \texttt{729.abc} comparing \texttt{refrate} against \texttt{refspeed}. While support coverage is nearly complete (\(99.9\%\)), the temporal phase structure reveals significant distributional skew (\(\mathrm{TVD}=86.2\%\), weighted accuracy \(31.8\%\)). High code coverage does not guarantee that execution time is spent in the same relative proportions.}
  \label{fig:abc-refrate-refspeed-bbv}
  \Description{BBV self-similarity plot comparing refrate and refspeed for 729.abc showing high coverage but large distributional divergence.}
\end{figure}

Finally, the refrate/refspeed comparison demonstrates that representativeness
gaps are not restricted to training inputs. For \texttt{729.abc}, \texttt{refrate}
covers \(99.986\%\) of \texttt{refspeed} dynamic BBV mass, \(0.999\) of branch
targets, \(0.998\) of callgraph edges, and \(0.999\) of CFG edges
(\autoref{tab:case-studies}, row 6). By conventional support coverage,
\texttt{refrate} appears to be an exhaustive superset. However, as shown in
\autoref{fig:abc-refrate-refspeed-bbv}, the shared behavior is weighted
differently. Weighted accuracy drops to \(31.8\%\), and Total Variation
Distance reaches \(86.2\%\). Phase intervals that dominate \texttt{refrate}
shrink to brief transients in \texttt{refspeed}, and vice versa. This case
uses the identical compiled rate binary image for both executions; the divergence
therefore arises from workload invocation and input scale rather than compiler
flags or binary layout. It
underscores that high coverage cannot serve as a proxy-quality certificate when
relative phase weights are heavily distorted.

\section{Related Work}
\label{sec:related-work}
We position our work against prior research in benchmark representativeness, phase sampling, dynamic binary instrumentation, and profile comparison.

\subsection{Proxy Representativeness}

Benchmarking suites serve as standard proxies for real-world application behavior across computer systems performance evaluation~\cite{cpu2017, cpu2026}. However, the empirical fidelity of benchmark proxies depends on input selection. Eeckhout et al.~\cite{input_datasets} established that benchmark input datasets alter dynamic execution paths, cache utilization, and microarchitectural bottlenecks. Subsequent performance-engineering research formalized workload similarity using statistical clustering, principal component analysis, and distance metrics over hardware performance counters to quantify benchmark redundancy and suite coverage~\cite{bm_similarity, server_wkld_distances, alberta_workloads}.

In standard benchmark suites, reduced or `train' inputs are routinely employed in place of full reference runs to reduce execution turnaround. Gove and Spracklen~\cite{cpu2006_reftrain, cpu2006_workshop} examined SPEC CPU2006, questioning whether training inputs faithfully exercise the same execution paths as reference workloads. While their investigation evaluated this discrepancy primarily through the lens of compiler feedback, our work formulates a broader, architecture-agnostic measurement contract. We extend beyond simple basic-block counts to capture multi-faceted control-flow distributions---including direction bias, history-conditioned correlation, indirect branch targets, dynamic callgraphs, and CFG edge topologies---applicable to any paired execution of an identical binary across distinct inputs or runtime configurations.

\subsection{Sampling, BBVs, and Phase Dynamics}

To mitigate the massive execution overhead of cycle-accurate architectural simulation, input-reduction initiatives such as MinneSPEC \cite{minnespec} developed truncated workloads, though subsequent evaluations demonstrated that synthetic downsizing frequently distorts algorithmic behavior. Rather than constructing artificial inputs, SimPoint~\cite{bbv_distribution, simpoints, simpoints3} popularized frequency vectors of basic blocks (BBVs) to divide dynamic instruction streams into fixed-length intervals, clustering them into recurrent behavioral phases to identify compact, statistically representative simulation windows. Complementing interval clustering, SMARTS~\cite{smarts} leveraged periodic statistical sampling to approximate full-length benchmark behavior with high statistical confidence.

In our framework, basic-block vectors serve an entirely distinct evaluative objective. Rather than selecting intra-run checkpoints for a single execution, we construct whole-run BBV frequency distributions and cross-execution recurrence maps to audit whether a candidate proxy reaches and proportionally weights the same dynamic phases as its reference target. While recurrence and self-similarity matrices have established utility across dynamical systems~\cite{recurrence_complex_systems}, acoustic analysis~\cite{music_viz, music_ss}, human motion tracking~\cite{action_ss}, and memory-trace visualization~\cite{mav_perf}, our workflow applies normalized Euclidean BBV distance matrices directly in the instruction domain. This representation cleanly distinguishes internal phase progression along the diagonal from cross-execution divergence in off-diagonal blocks.

\subsection{Dynamic Binary Instrumentation}

Dynamic binary instrumentation (DBI) engines provide non-invasive execution observability without requiring source access, debug symbols, or compiler modifications~\cite{valgrind}. Within the Valgrind ecosystem, tools such as Cachegrind~\cite{cachegrind_manual} couple binary instrumentation with simulated hardware models, estimating branch mispredictions through fixed-capacity two-bit saturating counter tables and branch-target buffers. However, microarchitectural simulation binds observational data to a specific, idealized hardware design, predicting machine-dependent failure rates rather than workload properties.

In contrast, \texttt{exp-branchprof} intentionally eschews hardware modeling. It captures raw, empirical frequency distributions of dynamic control-flow events across branches, histories, targets, and call sites. Preserving concrete event multisets enables symmetric, machine-independent comparisons between paired executions of unmodified binaries. Building upon the foundational insight of Yeh and Patt~\cite{branch_correlation_92, two_level_corr} that recent global history exposes branch correlation patterns, our tool records a lightweight 4-bit history context not to simulate a hardware predictor, but to quantify how local execution context reduces empirical outcome entropy between workload pairs.

\subsection{Structural Graph Recovery}

Binary analysis frameworks such as CFGgrind~\cite{cfggrind}, Dyninst \cite{dyninst_binary_code_not_easy}, and DCFG/PinPlay~\cite{dcfg_pinplay} emphasize structural recovery, reconstructing complete, topologically sound control-flow graphs from machine code for decompilation, program understanding, or binary rewriting. In contrast, we specifically tailor \texttt{exp-branchprof} for \textbf{paired statistical workload comparison}. Extending classical edge-frequency instrumentation concepts~\cite{ball_larus_profiling_1994}, our profiler aggregates address-keyed event multisets into compact mathematical rollups rather than maintaining explicit graph topologies with source attribution.

This architectural distinction yields two practical advantages for performance engineering. First, it minimizes runtime memory footprint and eliminates graph-traversal overhead during collection. Second, it enables closed-form statistical distance calculations such as total variation distance, event-mass coverage, and fan-out entropy. It also facilitates privacy-preserving summary exports, allowing industrial engineering teams to share diagnostic fidelity metrics for proprietary customer workloads without disclosing proprietary source code, symbol names, or instruction bytes. Furthermore, while binary similarity frameworks~\cite{codeflow, binary_similarity, cebin_similarity} compare static or dynamic graphs across varying compilation settings or architectures, our protocol strictly holds the binary fixed, isolating input-induced control-flow drift.

\subsection{Profiler Trust, Validation, and Optimization}

A primary requirement in performance engineering is profiler correctness and repeatability. Mytkowicz~\cite{mytkowicz2009} demonstrated that measurement bias can produce misleading conclusions in systems evaluation, while Weaver~\cite{weaver-hipeac2008} exposed the non-determinism and hardware counter inaccuracies common across modern processors. To establish empirical trust, our methodology incorporates exact-oracle semantic validation on deterministic, non-PIE AArch64 microbenchmarks, confirming that all recorded events match architectural ground truth down to individual CSV rows, complemented by operational bounds on run stability and memory overhead.

The challenge of comparing execution profiles has an extensive history. Information-theoretic profile blending was introduced by Savari~\cite{savari_combining_profiles_2000}, who applied relative entropy to reconcile disparate branch profiles. In early investigations, Wall~\cite{wall_predicting_profiles_1991} explored empirical profile estimation, while Wus~\cite{wu_static_branch_1994} devised static branch heuristics to predict path probabilities. Langdale~\cite{langdale_profile_usefulness_2003} demonstrated that \textbf{profile accuracy does not imply profile usefulness}. High similarity under a given profile metric does not guarantee proportional performance gains in downstream consumers because compiler heuristics, inlining thresholds, and register allocation mediate the outcome.

While a concurrent study focuses specifically on compiler optimizations and investigates how train-input fidelity gaps degrade downstream Profile-Guided Optimization across compilers and architectures~\cite{companion_study_cgo}, our framework intentionally decouples \textbf{workload representativeness measurement} from downstream optimizer heuristics. By treating multi-view control-flow fidelity as an objective diagnostic property of the workload pair, our methodology provides performance engineers with actionable visibility into proxy representativeness before any specific optimization or simulation pipeline is invoked.

\section{Conclusion}
\label{sec:conclusion}

A cheaper execution is not automatically a representative proxy for a longer
one. It can reach the same code while missing target event mass, changing
relative hotness, or altering the dynamic call and control-flow structure. This
paper presents a measurement contract to make those differences
explicit. The methodology compares two controlled executions of the same
binary using BBV phase and histogram distributions alongside five address-keyed
BranchProf profile families.

\texttt{exp-branchprof} records conditional directions and short-history
contexts, branch targets, callgraph edges, and CFG edges from unmodified
binaries. Exact-oracle tests validate its handling of conditional and inverted
exits, direct and indirect jumps, calls, returns, and process-separated output.
The SPEC cases show both high-fidelity controls and mismatches that would be
hidden by code reachability alone, including a callgraph mismatch despite nearly
complete branch-target coverage.

The intended output is evidence for an engineering decision, not a universal
proxy score or a model of a particular compiler or processor. The accompanying
artifact packages the pinned Valgrind patch, tests, collection scripts, and
aggregate rollups so that proprietary workload/proxy pairs can be measured
without returning raw PC-level profiles.

% Activate the BST control entry to show DOIs
%\bstctlcite{IEEEtran_bst_ctl}

% Add your BibTeX references in 90-citations.bib
%\bibliographystyle{IEEEtranS}
\bibliographystyle{ACM-Reference-Format}
\bibliography{90-citations}

%%% The arXiv version includes the supplementary appendix.
\ifextendedappendix
% Appendix material

\appendix

\section{Extended Metric Interpretation and Visualization Details}
\label{app:metric-interpretation}

This appendix records supplementary reading aids and derived quantities for the
metrics defined in \autoref{sec:comparison}.

\paragraph{BBV self-similarity plots.}
The self-similarity matrices use the normalized BBV distance \(D(i,j)\) defined
in \autoref{sec:comparison}. Square-root color normalization expands contrast
among small distances; the train/reference boundary is a separator rather than
a data row or column. The colorbar represents BBV distance, so smaller values
indicate more similar basic-block mixtures. The matrices preserve execution
order, but their dimensions are intervals defined by instruction count, not
wall-clock time.

For aggregate BBV histograms, cosine similarity is
\begin{equation}
  \mathrm{cosine}(T,R)
  = \frac{\sum_b c_T(b)c_R(b)}
  {\sqrt{\sum_b c_T(b)^2}\sqrt{\sum_b c_R(b)^2}}.
\end{equation}
It emphasizes alignment of the unnormalized histogram directions and is
insensitive to total event volume.

\paragraph{Entropy units and profile-specific interpretation.}
All entropy and entropy-error values use base-2 logarithms and are reported in
bits (shannons). A balanced binary branch has \(1\) bit of entropy, and \(k\)
equally likely targets have \(\log_2 k\) bits. Entropy reductions and errors
are in bits; relative reductions are unitless. For the profile families:
\begin{itemize}[leftmargin=10pt]
\item Conditional-branch entropy is taken/not-taken uncertainty at a static
branch PC. Zero bits means that only one outcome was observed; one bit means
the outcomes were evenly split.
\item Branch-history entropy is the remaining outcome uncertainty after
conditioning on the four-bit global-history bucket. Its reduction is the
uncertainty explained by that short history.
\item Branch-target entropy is uncertainty over the target PCs observed from a
branch or call site. A monomorphic site has zero bits; \(k\) equally likely
targets have \(\log_2 k\) bits.
\item Callgraph target entropy is callee-target uncertainty at a callsite, and
caller entropy is incoming-caller uncertainty at a callee.
\item CFG successor entropy is destination uncertainty from a source PC, and
predecessor entropy is incoming-source uncertainty at a destination PC.
\end{itemize}

\paragraph{Dynamic entropy rankings.}
For within-execution diagnosis, the artifact ranks a branch by
\(C_b H(b)\), its dynamic uncertainty contribution in bit-events, and by
\(C_b\Delta H(b)\), the dynamic uncertainty explained by four-bit history.
These rankings distinguish rare high-entropy branches from branches that
materially contribute to an execution's uncertainty. They are diagnostic
quantities, not additional proxy-fidelity criteria; \(\Delta H(b)/H(b)\), when
\(H(b)>0\), is the corresponding unitless fraction explained by history.

\section{Exact-Oracle Validation}
\label{app:semantic-validation}
\autoref{tab:semantic-validation} gives the complete expected-versus-observed
matrix for the deterministic, address-labelled exact-oracle microbenchmarks
summarized in \autoref{sec:validation}.

\begin{table}[hb]
\footnotesize
\centering
\setlength{\aboverulesep}{0pt}
\setlength{\belowrulesep}{0pt}
\setlength{\extrarowheight}{.75ex}
\caption{Exact-oracle semantic validation. Each expected count is checked
against the address-labelled CSV row emitted by BranchProf.}
\label{tab:semantic-validation}
\begin{tabularx}{\dimexpr\columnwidth - 8pt\relax}{@{}
  >{\hsize=0.90\hsize}L
  >{\hsize=1.70\hsize}L
  >{\hsize=0.40\hsize}C
  @{}}
\arrayrulecolor{lightgray!40}
\rowcolor{tablehead}
\thead[l]{\textbf{\textcolor{white}{Case}}} &
\thead[l]{\textbf{\textcolor{white}{Expected Dynamic Record}}} &
\thead[c]{\textbf{\textcolor{white}{Result}}} \\
\bottomrule
Conditional exit & branch taken/not-taken \(=6/4\); two CFG edges \(=6,4\) &
pass \\
\bottomrule
Inverted \texttt{Ist\_Exit} & branch taken/not-taken \(=7/3\); CFG edge \(=10\) &
pass \\
\bottomrule
Direct call and return & callgraph and call CFG edge \(=11\); return CFG and
target edge \(=11\) & pass \\
\bottomrule
\texttt{Ijk\_Boring} direct jump & CFG edge \(=13\) & pass \\
\bottomrule
Indirect jump & two CFG and target edges \(=5,4\) & pass \\
\bottomrule
\texttt{fork()} file separation & two complete profile sets; branch
taken/not-taken \(=3/4\) and \(6/2\) & pass \\
\bottomrule
Two pthreads & aggregate branch and CFG counts \(=4/4\); independent
history rows \texttt{TTTT} (taken) and \texttt{NNNN} (untaken) & pass \\
\bottomrule
\end{tabularx}
\par
\setlength{\aboverulesep}{0.6ex}
\setlength{\belowrulesep}{0.9ex}
\setlength{\extrarowheight}{0pt}
\end{table}

\section{Artifact Appendix}
\label{app:artifact}

\paragraph{Abstract.}
The artifact provides the implementation, deterministic validation suites, raw
illustrated-case profiles, repeatability data, aggregate corpus summaries, and
analysis scripts used in this paper. It regenerates the numerical inputs for
the operational-cost and case-study tables, the BBV recurrence plots, and the
repeated-run stability results. It also provides configuration stanzas for
collecting new profiles with a separately licensed SPEC CPU installation.

\paragraph{Artifact checklist.}
\begin{itemize}[leftmargin=10pt]
\item \textbf{Program:} Source patches for Valgrind
  \texttt{exp-branchprof}, AArch64 and x86-64 exact-oracle microbenchmarks,
  and a five-profile-family smoke test.
\item \textbf{Data:} Raw BranchProf and BBV profiles for the six illustrated
  cases, 90 repeated-run profiles, all-suite aggregate rollups, and sanitized
  matched native/instrumented runtime samples. SPEC CPU source, binaries, and
  inputs are not redistributed.
\item \textbf{Requirements:} The data-reproduction path requires Linux and
  Python 3.10 or newer. Rendering also requires Matplotlib; rebuilding BBV
  matrices requires a C++20 compiler. New collection and the exact semantic
  tests require patched Valgrind on AArch64 or x86-64. Root privileges are not
  required.
\item \textbf{Metrics:} Support and event-mass coverage, total variation
  distance, taken-probability MAE, entropy in bits (shannons), Jensen--Shannon
  divergence, and BBV cosine/distance measures.
\item \textbf{Output:} CSV tables, six PNG recurrence plots, raw-rollup
  reports, repeatability summaries, and terminal semantic-test assertions.
\item \textbf{Variation:} Generated tables and repeatability reports are
  checked byte-for-byte. Rebuilt BBV matrices allow maximum absolute difference
  \(10^{-6}\), accounting for parallel floating-point reduction.
\end{itemize}

\paragraph{Access and workflow.}
The artifact is permanently archived at
\url{https://doi.org/10.5281/zenodo.24160599} \cite{artifact}. From the extracted artifact
root, the fast deterministic path is:
\begin{verbatim}
python3 artifact.py reproduce-and-verify --output result
\end{verbatim}
It verifies 12 deterministic outputs without SPEC CPU, a compiler, Valgrind,
network access, or root access. The complete bundled-data workflow is:
\begin{verbatim}
python3 artifact.py all --output result
\end{verbatim}
It additionally renders six plots, recomputes six raw case comparisons,
verifies repeatability, and rebuilds six BBV matrices. The artifact's
\texttt{README.md} and \texttt{GETTING\_STARTED.md} give prerequisites,
expected outputs, a portable container path, and a staged guide for new users.

\paragraph{Resources and scope.}
The compressed archive is approximately 111\,MB and expands to approximately
0.9\,GB; reserve 2\,GB for the full workflow. On a 32-core AArch64
evaluation host, the complete workflow takes about ten minutes; on a tested
40-vCPU x86-64 host, it takes about four minutes. The address-keyed comparisons
require the same binary image and layout for both executions and characterize
control-flow agreement; they do not predict a specific compiler transformation
or microarchitectural effect.

\fi
\end{document}